\documentclass[epj]{svjour}  
\usepackage{booktabs}
\usepackage{multirow}
\usepackage[draft]{microtype}
\usepackage{diagbox}
\usepackage{cite}
\usepackage[final]{graphicx}
\usepackage[export]{adjustbox}
\usepackage{threeparttable}
\usepackage{textcomp}
\usepackage{url}
\usepackage{subfig}
\usepackage{amsmath}
\usepackage{nicefrac}
\usepackage{soul}
\usepackage{hyphenat}
\usepackage[shortcuts]{extdash} 
\usepackage[british]{babel}
\usepackage{etoolbox}
\usepackage{hyperref}

\usepackage{fetamont}
\usepackage[T1]{fontenc}
\usepackage{rotating}
\usepackage{relsize}
\newcommand*{\delfinfont}{\fontfamily{fetamont}\ffmfamily\bfseries\selectfont}
\newcommand{\delfin}{\textsc{{\delfinfont DE{\smaller$\ell{}$}FIN}}}

\newcommand{\nubarA}{$\bar{\nu}$(A)}
\newcommand{\nubar}{$\bar{\nu}$}
\newcommand{\Eexc}{E$^{\ast}$}
\newcommand{\EexcA}{E$^{\ast}$(A)}
\newcommand{\Erot}{E$_{\mathrm{rot}}$}

\usepackage{tikz}

\DeclareRobustCommand{\isotope}[2]{\textsuperscript{\tiny{#2}}#1}
\usepackage[inter-unit-product=\ensuremath{{}\cdot{}},load=physical,per-mode=symbol,detect-shape=true,separate-uncertainty,multi-part-units = brackets,product-units = brackets-power]{siunitx}

\begin{document}
%
%\title{\delfin{} -- a TALYS-based tool for the comparison of fission models}
%\title{\delfin{} -- a TALYS-based tool for the comparison of fission fragment evaporation models}
\title{Disentangling fission fragment evaporation models using TALYS}
%\subtitle{Do you have a subtitle?\\ If so, write it here}

\author{Andreas Solders
 \and Andrea Mattera \and Ali Al-Adili \and Mattias Lantz  \and Vasileios Rakopoulos \and Stephan Pomp} 
 
\mail{andreas.solders@physics.uu.se}

\institute{Uppsala University, BOX 516, 75120 Uppsala, Sweden}

%\email{e-mail: andreas.solders@physics.uu.se}
%\thankstext{e1}{e-mail: andreas.solders@physics.uu.se}
%\thankstext[e1]{fauthor@example.com}
% etc
% \thanks is optional - remove next line if not needed
%\thanks{\emph{Present address:} Insert the address here if needed}%
                    % Do not remove
%
%\offprints{}          % Insert a name or remove this line
%

%
\date{Received: date / Revised version: date}
% The correct dates will be entered by Springer
%

\abstract{
Several computer codes based on phenomenological models are being developed with the aim of obtaining fission observables, such as neutron and gamma multiplicities and product yields. Key points in these calculations, which are handled differently by the various codes, are the sharing of the total excitation energy between the fragments and the generation of angular momenta. After the initial states of the fragments is set, the de-excitation through the emission of neutrons and photons can be model. However, many models also include tunable parameters that are used to make the calculations conform with literature data. As a result, it could be possible to obtain good agreement with experimental results even with questionable assumptions on the initial conditions. %The calculations typically start with the excited fragments that are consecutively de-excited through the emission of neutrons and photons. However, the staring point, the sharing of the total excitation energy between the fragments and the generation of fragments angular momenta, is a key point which is handled differently by the various codes. Moreover, even with questionable assumptions on the initial conditions it could be possible to get good agreement with experimental data through tuning of the de-excitation process.
To test the assumptions made by different fission models the code \delfin{} has been developed. \delfin{} starts with the fission fragments as generated by the models and calculates the average neutron emission as a function of mass in a transparent and coherent way using the nuclear reaction code TALYS. Hence, the probabilities of neutron emission in competition with $\gamma$ de-excitation comes from parameters of the TALYS code that have not been optimised. The results are then compared looking at general trends and at the difference to what is obtained with the stand-alone versions of the codes.
In this study, the output of \delfin{} for five of the most commonly used fission codes is compared to the results of the stand-alone versions, and to experimental data, for the reactions \isotope{U}{235}(n$_{\mathrm{th}}$,f), \isotope{Pu}{239}(n$_{\mathrm{th}}$,f) and \isotope{Cf}{252}(sf).
\PACS{
      {21.60.-n}{Nuclear structure models and methods} \and
      {24.75.+i}{General properties of fission}   \and
      {25.85.Ca}{Spontaneous fission} \and
      {25.85.Ec}{Neutron-induced fission}
     } % end of PACS codes
} %end of abstract

\maketitle
\section{Introduction}
\label{intro}
Almost 80 years after its discovery, a complete theory for the fission process is still missing. 
In the past years, extensive efforts have been made to improve the description of fission and several models are under development.

The ambitious endeavour of \textit{ab initio} models is still far from the needs of, for example, nuclear data applications, both in terms of precision and computing time.
Phenomenological models on the other hand, though less rigorous, often use existing nuclear data as input or as a way to constrain the values of model parameters.
As a consequence, they  are generally able to reproduce experimental data, but their power to extrapolate to  fissioning systems where data are scarce may be limited.

Over the years, many codes have been developed to utilize such phenomenological models. The very first attempt, the Madland-Nix Los Alamos Model \cite{madland1982new}, later extended and included into Vladuca and Tudora's PbP model \cite{vladuca2001prompt,tudora2006experimental}, was able to describe integral fission properties. More lately, Monte Carlo codes such as GEF \cite{schmidt2010general}, \texttt{FREYA} \cite{randrup2009calculation,vogt2009event}, CGMF \cite{kawano2010monte,talou2011advanced,becker2013monte} and FIFRELIN \cite{fifrelin,litaize2010investigation} have been developed.  These codes simulate the de-excitation of fission fragments on an event-by-event basis and can therefore account for fluctuations of different observables as well as correlations between, e.g., neutron and gamma multiplicities.

Regardless of whether it is based on the Monte Carlo method or on deterministic calculations, the main components of the studied fission models are essentially the same.
As a starting point the available energy must be estimated, based on the $Q$ value for a given mass split of the Fission Fragment (FF) pair.
This can be done relying directly on experimental data (\textit{e.g.}, using the total kinetic energy as a function of FF, $TKE$(A,Z), as most of the models listed above do) 
or, as in the GEF code, using a phenomenological description of the fission event to calculate the total excitation energy $TXE$ \cite{schmidt2014general}.
%barrier height whose parameters are then fit to experimental data.

From the $TKE$ the $TXE$ can be obtained (or the other way around), using the relation:
\begin{equation}
\begin{split}
 E_{\mathrm{tot}} &= TXE + TKE\\
		  &= E_n + S^{\mathrm{CN}}_n + Q ,
\label{eqn:Etot-general}
\end{split}
\end{equation}
where $E_\mathrm{n}$ is the kinetic energy of the incoming neutron, $S^{\mathrm{CN}}_\mathrm{n}$ the neutron separation energy of the compound nucleus (both of which are zero in the case of spontaneous fission), and $Q$ the energy released in the reaction.

The available excitation energy is subsequently shared between the two FFs. 
This step is possibly the most debated upon \cite{tudora2015comparing,andreyev2017nuclear,tudora2017comprehensive} and models often introduce various corrections to the underlying physical assumptions in order to reproduce experimental data.

% Once a code manages to bring the experimental data (like, e.g., nubar(A)) into accordance with assumed initial state of the fission fragment right after scission, it is the task of fission models to describe the path from compound nucleus formation until scission with fragments in the states that de-exciation models derived (or deduced). 

The main focus of the models is the definition of the properties of the FFs after scission. However, once the physical parameters of the fission event are fixed, fission observables, such as the prompt neutron multiplicity or the total gamma energy, have to be derived. To do so the authors include some treatment of the de-excitation of the FFs. Fission quantities extracted from these codes thus appear as the convolution of these two steps.

The aim of this work is to compare the basic assumptions made by different fission fragment evaporation models by eliminating any variability in the way the final observables are extracted. 
This can be achieved using the quantities defined \emph{right after scission} by the models and introducing a transparent and coherent way of handling the fragment de-excitation. 

\section{Method}
\label{M&M}

To carry out the de-excitation of the FF, and to extract the desired observables, the TALYS reaction code \cite{talys} was used. 
TALYS was chosen not only for the wide applicability of its models, but also for the fact that it is available, well documented and open source. 
Furthermore, TALYS allows the user to have full control on the reaction parameters. 

The preparation of the TALYS inputs from the distributions of FF excitation energies and, if available, the angular momentum, as well as the subsequent extraction of the final observables, have been bundled in a single code: \delfin{} (De-Excitation of FIssion fragmeNts) \cite{mattera2017methodology}. The excitation energies can either be averaged over Z for each mass (\EexcA), or on a nuclide basis (\Eexc(A,Z)). In this study the de-excitation is  performed using TALYS\=/1.8 and the observable chosen for the comparison is the average prompt neutron multiplicity as a function of fragment mass (\nubarA{}).

%%%%%%%% Next paragraph need to be changed %%%%%%%%%%
%%%%%%%%%%%%%%%%%%%%%%%%%OLD%%%%%%%%%%%%%%%%%%%%%%%%%
% The description and the validation of the methodology behind \delfin{}, as well as a first comparison of the GEF, PbP and \texttt{FREYA} models was reported in ref.~\cite{mattera2017methodology}.
% In this work we discuss an extension of our study to include CGMF (for \isotope{Cf}{252}(SF)) and FIFRELIN (for \isotope{Cf}{252}(SF), \isotope{U}{235}(n$_{\mathrm{th}}$,f) and \isotope{U}{239}(n$_{\mathrm{th}}$,f)).
%%%%%%%%%%%%%%%%%%%%%%%%%%%%%%%%%%%%%%%%%%%%%%%%%%%%%
%%%%%%%%%%%%%%%%%%%%%%%%%NEW%%%%%%%%%%%%%%%%%%%%%%%%%
The principle followed in the development of \delfin{} is transparency; as far as possible, no arbitrary assumptions are made on the quantities needed for the calculations. 
The process is completely reproducible and the default values of the physical parameters in the TALYS code were not modified for this study.
If a value needed for the calculation is not provided by the models themselves (\textit{e.g.}, FF yields, $Z$($A$) distributions, \textit{etc.}) we turn to well-known systematics \cite{wahl1988nuclear}.

Not all codes for the simulation of fission observables provide the same information on the FFs.
Deterministic codes often only give quantities averaged over mass-chains, but even most Monte Carlo codes do not provide an output where the event-by-event information is readily available.
Hence, it was necessary to investigate to what extent the input could be simplified (\textit{i.e.}, how little information would be required from the model) before the results would start to deviate considerably from the full calculation.
The evaluation of the method was made using the GEF code, as it provides extensive information on the FFs on an event-by-event basis. This evaluation showed that the use of the average excitation energy for the isobaric chains (\EexcA) gives results that are in very good agreement with those obtained using the full distribution of excitation energies (\Eexc(A,Z)), provided that several FFs in the mass chain, to which is assigned the average excitation energy, are included in the calculation \cite{matteraPhDthesis,mattera2017methodology}.

\begin{figure}
\centering
\includegraphics[width=\columnwidth]{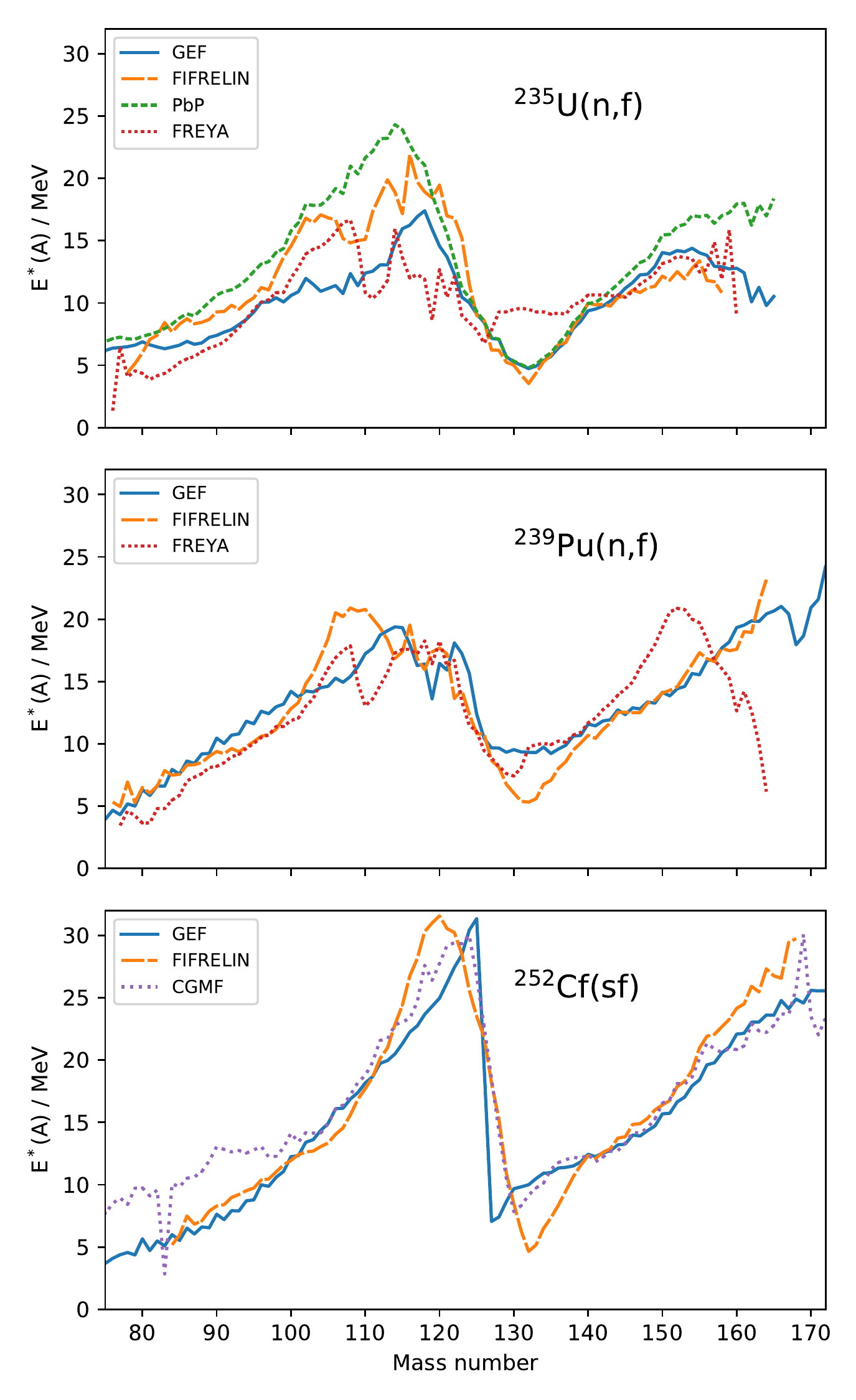}
\caption{Fission fragment excitation energy as a function of fragment mass before neutron emission, as obtained from each model code, for the three reactions studied.}
\label{fig:exc_ene}
\end{figure}

The results presented here concern five of the most commonly used models for the description of fission observables: GEF, FIFRELIN, \texttt{FREYA}, PbP and CGMF.
The codes were tested using three well studied fission reactions, \isotope{U}{235}(n$_{\mathrm{th}}$,f), \isotope{Pu}{239}(n$_{\mathrm{th}}$,f) and \isotope{Cf}{252}(sf), and the calculations were made using the average excitation energies as extracted from the respective code, see figure~\ref{fig:exc_ene}.

Neither \texttt{FREYA} nor FIFRELIN nor PbP provide FF yields. Therefore, these calculations had to be performed using mass averaged yield distributions (Y(A)) extracted from the Wahl systematics \cite{wahl1988nuclear}.
The GEF code, on the other hand, does provide information on the independent fragments yields (Y(A,Z)). However, for consistency, the Wahl systematics was used also for the GEF calculations. This choice does not affect the final result~\cite{mattera2017methodology}.
CGMF, finally, provide mass averaged FF yields. For comparison, the calculations in \delfin{}  were performed both for the Wahl systematics and for the the yields provided by CGMF. The latter is then labeled ``DELFIN (FY CGMF)''.

It should be pointed out that since the \delfin{} code detaches the fission models from the FF de-excitation, and processes this through TALYS, the probabilities of neutron emission in competition with $\gamma$ de-excitation come from parameters in the TALYS code.
These parameters have not been optimized and it is not in the scope of this work to try to reproduce experimental data. % (\textit{e.g.}, with a Total Monte Carlo approach \cite{TMC}).
Hence, the results should be compared looking at general trends and at the difference to what is obtained with the stand-alone versions of the codes, rather than at specific deviations from experimental data.
% (in the \nubarA{} distributions in fig.~\ref{fig:nubar-cf252sf}-\ref{fig:nubar-pu239+nth}, or --- later --- in the total \nubar)

%%%%%%%%%%%%%%%%%%%%%%%%%%%%%%%%%%%%%%%%%%%%%%%%%%%%%

\section{Results}
\label{Results}

In each case the result obtained using \delfin{} was compared with the results from the stand-alone codes, as well as with well established experimental or compiled data. A summary of the results can be found in table~\ref{tab:nubar} where the total neutron multiplicity averaged overall masses of the FF mass distribution, \nubar, as well as the \nubar{} of the light and heavy fragments, are tabulated for all codes and reactions studied. The \nubar{} of the respective sides are derived by folding the calculated \nubarA{} with well-known mass distributions \cite{geltenbort1986precision, gook2014cf252}.
As reference, the total neutron multiplicities from the ENDF/B-VII.1  database \cite{endf} are also listed.

\begin{table*}
%\centering
%\vspace{.8cm}
\caption{Total average prompt neutron multiplicity of the different reactions under study obtained using the respective codes as stand-alone (s.a) and with \delfin{}. The values are reported as $\bar{\nu} ^{\bar{\nu}_\mathrm{LF}} _{\bar{\nu}_\mathrm{HF}}$, where $\bar{\nu}_\mathrm{LF}$ is the prompt neutron multiplicity averaged over the light mass region and $\bar{\nu}_\mathrm{HF}$ is the prompt neutron multiplicity averaged over the heavy mass region.}
\label{tab:nubar}
\resizebox{\linewidth}{!}{%
\begin{tabular}{l|c|cc|cc|cc|ccc|ccc}
\toprule\addlinespace[0.5em]
\multirow{3}{*}{Reaction} & \multirow{3}{*}{\begin{tabular}[x]{@{}c@{}}ENDF\\B-VII.1\end{tabular}}&\multicolumn{2}{c}{GEF} & \multicolumn{2}{c}{\texttt{FREYA}} &\multicolumn{2}{c}{PbP} &\multicolumn{3}{c}{FIFRELIN} & \multicolumn{3}{c}{CGMF}\\
 & & \multirow{2}{*}{s.a.}  & \multirow{2}{*}{\delfin} & \multirow{2}{*}{sa.} & \multirow{2}{*}{\delfin} & \multirow{2}{*}{s.a.} & \multirow{2}{*}{\delfin}  & \multirow{2}{*}{s.a.} & \multirow{2}{*}{\delfin} & \delfin & \multirow{2}{*}{s.a.} & \multirow{2}{*}{\delfin} &\delfin \\
 & & & & & & & & & & (w/o \Erot) & & & (FY CGMF) \\
 \noalign{\smallskip}\hline
 %\midrule
 \addlinespace[0.5em]
$^{235}$U(n,f) & 2.44 &2.48 $^{1.29}_{1.19}$ &2.52 $^{1.39}_{1.14}$&2.54 $^{1.29}_{1.25}$&2.45 $^{1.19}_{1.26}$&2.45 $^{1.36}_{1.09}$&2.80 $^{1.68}_{1.12}$&2.39 $^{1.40}_{0.99}$&2.93 $^{1.73}_{1.20}$&2.47 $^{1.48}_{0.99}$& --- & --- & ---\\\addlinespace[0.5em]
%2.51 $^{1.36}_{1.14}$& ?? &  2.40 $^{1.12}_{1.28}$ & ?? & 2.72 $^{1.61}_{1.11}$ &  & 2.72 $^{1.61}_{1.11}$& 2.40 $^{1.41}_{0.99}$ & 2.20 $^{1.30}_{0.90}$ & -- & -- \\ \addlinespace[0.5em]
$^{239}$Pu(n,f) & 2.87 & 2.84 $^{1.49}_{1.35}$&2.90 $^{1.60}_{1.30}$&2.94 $^{1.56}_{1.38}$&3.02 $^{1.55}_{1.47}$&---&---&2.83 $^{1.82}_{1.01}$&3.27 $^{1.97}_{1.30}$&2.83 $^{1.74}_{1.09}$&---&---&---\\ \addlinespace[0.5em]
%2.90 $^{1.56}_{1.34}$ & ?? $^{?}_{?}$ & 3.01 $^{1.46}_{1.55}$ & ?? $^{?}_{?}$ & -- & -- &3.22 $^{1.86}_{1.36}$ & ?? & 2.78 $^{1.73}_{1.05}$& -- & -- \\ \addlinespace[0.5em]
$^{252}$Cf(SF) & 3.76 & 3.75 $^{1.93}_{1.81}$&3.91 $^{2.06}_{1.85}$&---&---&---&---&3.69 $^{1.94}_{1.75}$&4.36 $^{2.35}_{2.00}$& 3.94 $^{2.15}_{1.79}$&3.75 $^{2.12}_{1.63}$&4.08 $^{2.24}_{1.83}$&4.09 $^{2.27}_{1.82}$\\
%3.92 $^{2.06}_{1.86}$ & ?? $^{?}_{?}$ & -- & -- & -- & -- & 4.37 $^{2.39}_{1.98}$ & ??$^{?}_{?}$ & 3.70 $^{1.96}_{1.73}$ & 4.11 $^{2.29}_{1.81}$ & 3.77 $^{2.13}_{1.64}$ \\ \addlinespace[0.2em]
\noalign{\smallskip}\hline
%\bottomrule

\end{tabular}}
\end{table*}

\subsection{\isotope{U}{235}(n$_{\mathrm{th}}$,f)}

\begin{figure*}
\includegraphics[width=\textwidth]{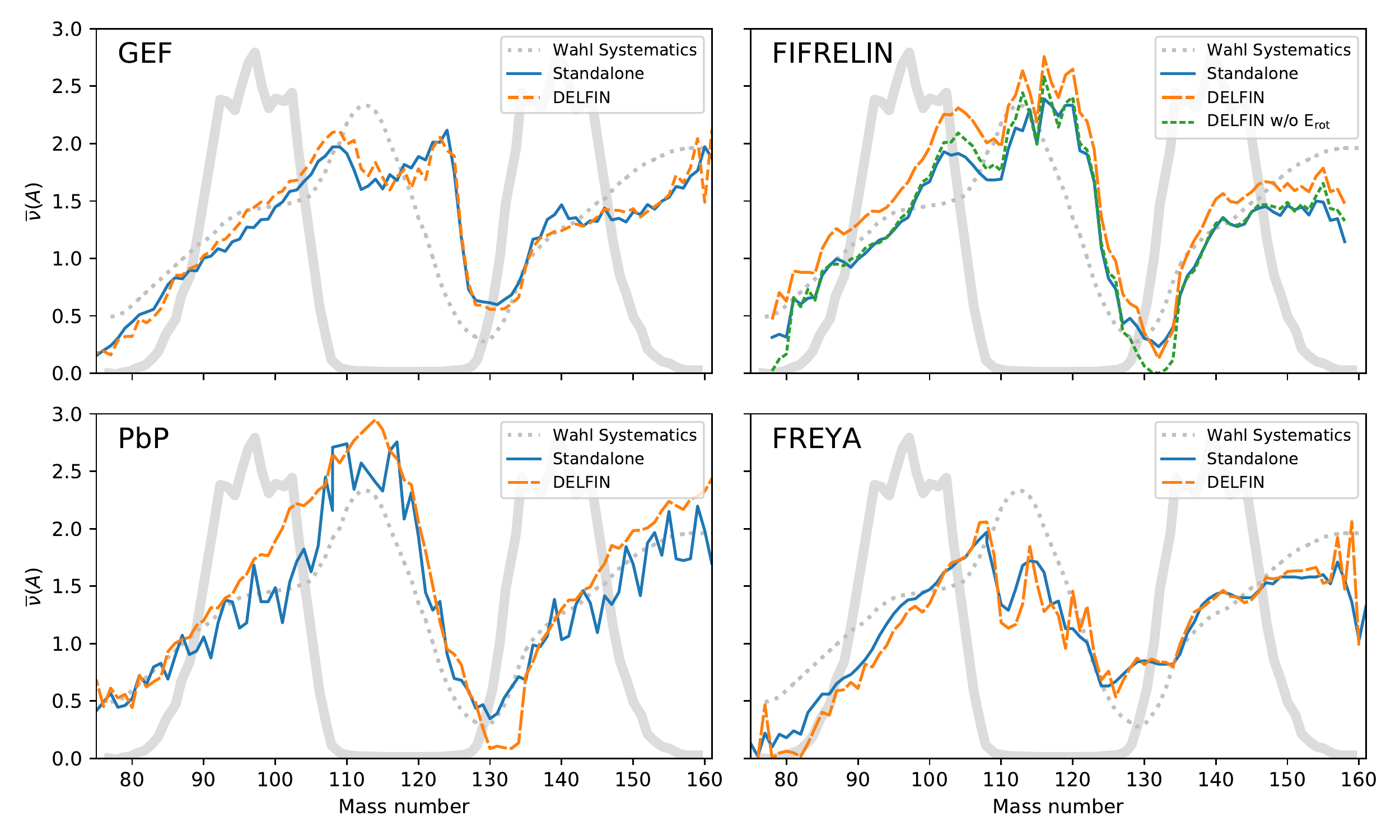}
\caption{\nubarA{} distributions from the \isotope{U}{235}(n$_\mathrm{th}$,f) reaction, as obtained by \delfin{} and the stand-alone codes using input from GEF, PbP, FIFRELIN and FREYA. The mass yields reported by Geltenbort \emph{et. al.}\cite{geltenbort1986precision}, that were used to calculate the values of \nubar{}, are plotted in the background.}
\label{fig:nubar-u235+nth}       % Give a unique label
\end{figure*}

The thermal neutron-induced fission of \isotope{U}{235} was studied using \delfin{} with \texttt{FREYA}, PbP, FIFRELIN and GEF and the results are shown in figure~\ref{fig:nubar-u235+nth}.
The $\bar{\nu}$(A) distributions from \delfin{} are compared with the Wahl systematics \cite{wahl1988nuclear} and with the \nubarA{} from the stand-alone codes.

Mass yields reported by Geltenbort \emph{et. al.}\cite{geltenbort1986precision} were used to calculate the total \nubar{} and the \nubar{} of the light and heavy fragments, respectively (see table~\ref{tab:nubar}). These yields are also plotted behind the \nubarA{} curves in figure~\ref{fig:nubar-u235+nth}.

\subsection{\isotope{Pu}{239}(n$_{\mathrm{th}}$,f)}

\begin{figure}
\centering
\includegraphics[width=\columnwidth]{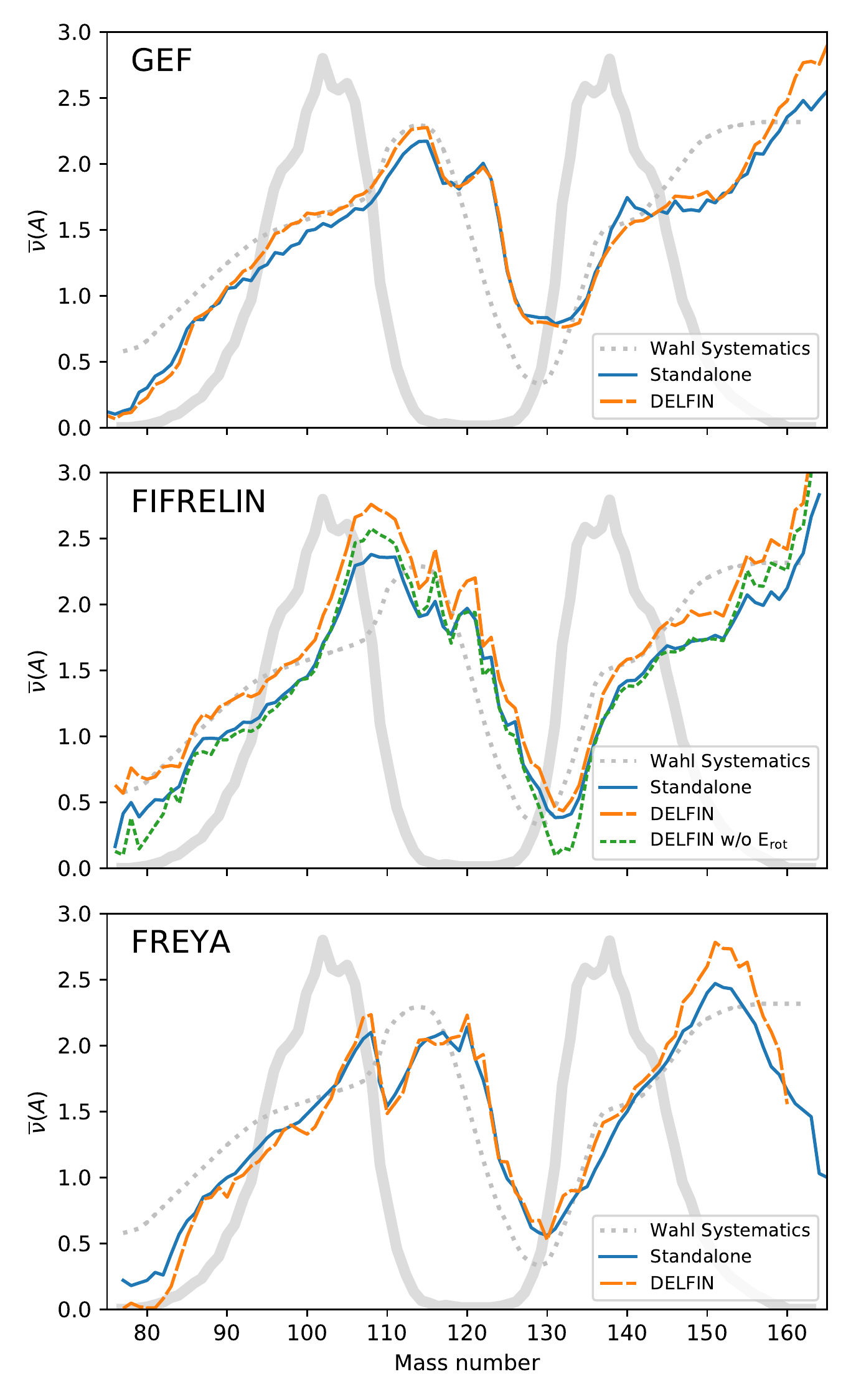}
\caption{\nubarA{} distributions for the \isotope{Pu}{239}(n$_\mathrm{th}$,f) reaction, as obtained by \delfin{} and the stand-alone codes using input from GEF, FIFRELIN and FREYA. The mass yields reported by Geltenbort \emph{et. al.}\cite{geltenbort1986precision}, that were used to calculate the values of \nubar{}, are plotted in the background.}
\label{fig:nubar-pu239+nth}
\end{figure}

The results for thermal neutron-induced fission of \isotope{Pu}{239} from the GEF, \texttt{FREYA} and FIFRELIN codes are shown in figure~\ref{fig:nubar-pu239+nth}, where the calculations with \delfin{} and the stand-alone codes are compared to Wahl systematics  \cite{wahl1988nuclear}. 

Mass yields reported by Geltenbort \emph{et. al.}\cite{geltenbort1986precision} are plotted behind the \nubarA{} curves in figure~\ref{fig:nubar-pu239+nth} and were used to calculate the \nubar{} vales in table~\ref{tab:nubar}.

%\begin{figure}
%\centering
%\includegraphics[width=0.99\columnwidth]{}
%\caption[Study of the effect of \Erot{} on the \nubarA{} distributions of \delfin{} with FIFRELIN for the \isotope{U}{235}(n$_\mathrm{th}$,f) reaction.]{Study of the effect of \Erot{} on the \nubarA{} distributions of \delfin{} with FIFRELIN for the \isotope{U}{235}(n$_\mathrm{th}$,f) reaction.}
%\label{fig:nubar-U235t-FIFRELIN+SA}
%\end{figure}

%\begin{figure*}
%\includegraphics[width=\textwidth]{Pu239}
%\caption{\nubarA{} distributions for the \isotope{Pu}{239}(n$_\mathrm{th}$,f) reaction as obtained from \delfin{} using input from GEF, FIFRELIN and FREYA.}
%\label{fig:nubar-pu239+nth}       % Give a unique label
%\end{figure*}

\subsection{\isotope{Cf}{252}(sf)}

\begin{figure}
\centering
\includegraphics[width=\columnwidth]{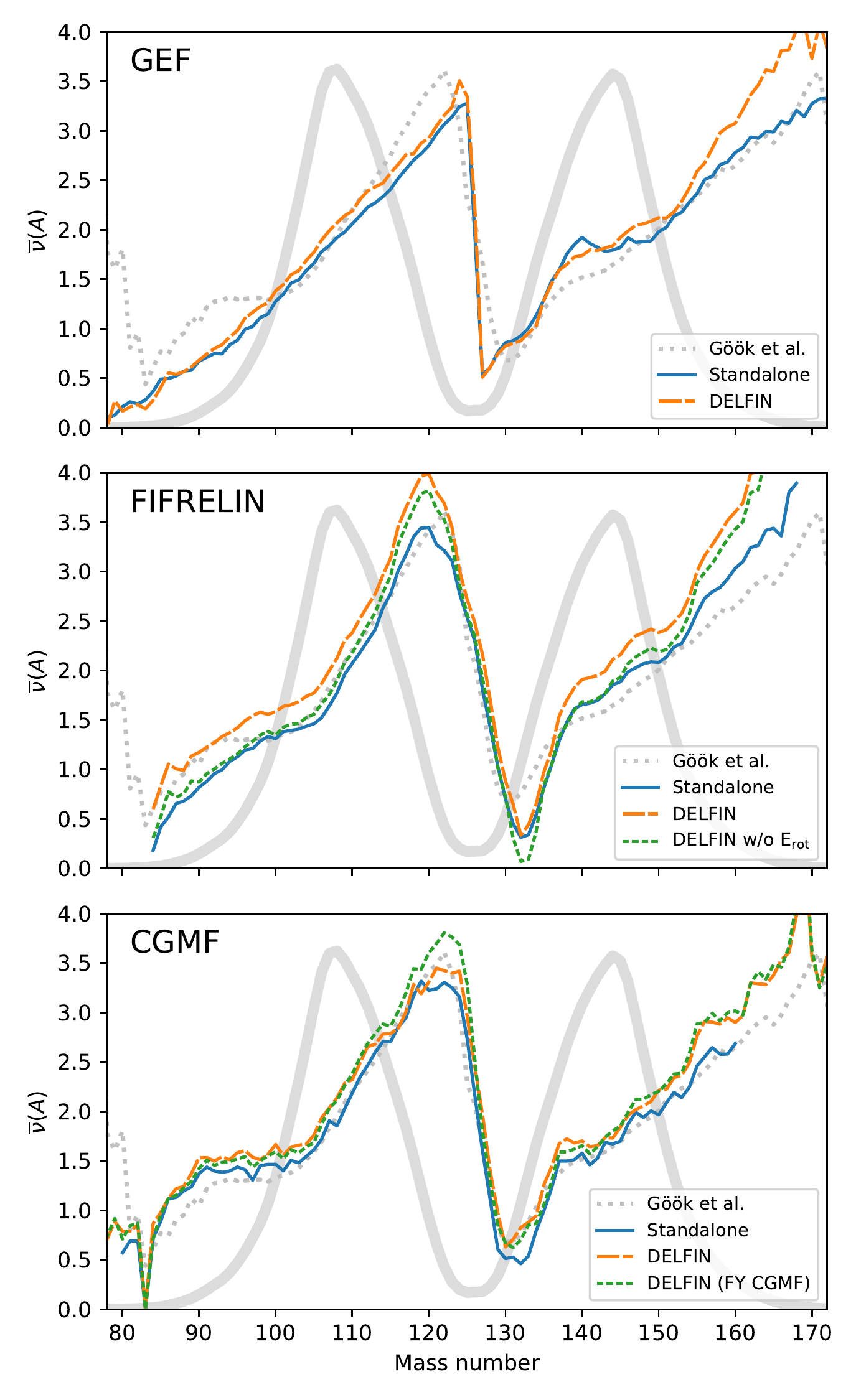}
\caption{\nubarA{} distributions for the \isotope{Cf}{252}(sf) reaction, as obtained by \delfin{} and the stand-alone codes using input from GEF, FIFRELIN and CGMF. The mass yields reported by G\"o\"ok \emph{et al.}\cite{gook2014cf252}, that were used to calculate the values of \nubar{}, are plotted in the background.}
\label{fig:nubar-cf252sf}
\end{figure}

The \nubarA{} for spontaneous fission of \isotope{Cf}{252} is shown in figure~\ref{fig:nubar-cf252sf} for \delfin{} with input from GEF, FIFRELIN and CGMF. The results are compared to the stand-alone codes and to experimental data from G\"o\"ok \emph{et al.}\cite{gook2014cf252}. 

The mass yields from G\"o\"ok \emph{et al.}\cite{gook2014cf252} are shaded behind the \nubarA{} curves and were used to calculate the \nubar{} vales in table~\ref{tab:nubar}.

%\begin{figure*}
%\includegraphics[width=\textwidth]{Cf252sf+SA}
%\caption{\nubarA{} distributions from \isotope{Cf}{252}(SF) as obtained from \delfin{} using input from GEF, CGMF and FIFRELIN.}
%\label{fig:nubar-cf252sf}       % Give a unique label
%\end{figure*}

\section{Discussion}

A general observation can be made, comparing \nubarA{} of the three different reactions (figs.~\ref{fig:nubar-u235+nth}~-~\ref{fig:nubar-cf252sf}). Over all, the differences between the model predictions are smaller for the \isotope{Cf}{252} reaction compared to the two others. The \isotope{Cf}{252}(sf) reaction is also the most studied and the one with the lowest uncertainties of the experimental data. Also notable is that within the same reaction, regions of low yield, where the uncertainties of the measurements are larger, show larger disagreements.
This demonstrates the close relationship between the development of the model codes and experimental data, which not only advocates for new and more precise measurements but also should be seen as a warning that it is risky to rely on models whenever data are scarce.

The region around double-magicity, at $A$~\num{= 132}, seems to be difficult to model.
Most often it is here that the calculated \nubarA{} distributions show their minimum, a few mass numbers above what is observed in the experimental data.
Discrepancies between the standalone calculations and the \delfin{} results, as well as the literature data, are in some cases also found around the mass of the complementary fragment ($A=A_{CN} - 132$).

%Below follow some observations that can be made in relation to the respective code.

% we extended the work presented in ref.~\cite{mattera2017methodology} to two additional fission codes: FIFRELIN (for all the studied reactions) and CGMF (for \isotope{Cf}{252}(SF)). 

% By doing this, we have reached a better understanding of what are the potentiality and the limits of this validation technique. 

%\begin{table}
%\caption{Please write your table caption here}
%\label{tab:1}       % Give a unique label
% For LaTeX tables use
%\begin{tabular}{lll}
%\hline\noalign{\smallskip}
%first & second & third  \\
%\noalign{\smallskip}\hline\noalign{\smallskip}
%number & number & number \\
%number & number & number \\
%\noalign{\smallskip}\hline
%\end{tabular}
% Or use
%\vspace*{5cm}  % with the correct table height
%\end{table}

%%% New paragraph from thesis %%%
\subsection{GEF}

The hump observed in the GEF stand-alone \nubarA{} of \isotope{U}{235} (fig.~\ref{fig:nubar-u235+nth}), \isotope{Pu}{239} (fig.~\ref{fig:nubar-pu239+nth}) and \isotope{Cf}{252} (fig.~\ref{fig:nubar-cf252sf}) around mass $A$~=~140 is not found in experimental data.
Neither is it reproduced when the same excitation energy distributions are processed in \delfin. 
This points towards an origin either from the de-excitation process of GEF or some mass dependent tuning parameter, rather than a consequence of the partition of the excitation energy (\Eexc{}) at scission.

The same is not true for the double-humped structure around mass 110 to 125 for \isotope{U}{235} and \isotope{Pu}{239}, that appears to be a direct consequence of the \Eexc{} distribution.
This shape, present in most of the models but at different positions, is not visible in experimental data \cite{PhysRevC.98.044615}. However, the experimental uncertainties in this region are high due to the low yields.

In general, except for these two regions that are worth exploring further, the agreement between \delfin{} and the stand-alone version of GEF is good.
For some of the reactions one can notice a slight over-estimation of the \nubarA{} from \delfin{} around A~=~100 and for A~>~155.
This could partly be due to the energy sorting mechanism built into GEF \cite{schmidt2010energysorting}. However, it could also be the result of the choice of parameters in TALYS, and cannot undoubtedly be attributed to features of the GEF model.

The agreement between \delfin{} and GEF shows that the Weisskopf evaporation model used by GEF is sufficient to reproduce the result of the more sophisticated Hauser-Feshbach model used in TALYS, at least for the reactions in this study.
It is also worth noting that the rotational energy (\Erot{}) of the fragments has been subtracted from the excitation energy provided by GEF \cite{mattera2017methodology}.
This suggests that GEF is able to derive a good estimate of the rotational energy which should provide a solid basis for the estimate of the $\gamma$ de-excitation with non-statistical photons, responsible for most of the angular momentum loss.

%%%%%%%%%%%%%%%%%%%%%%%%%%%%%%

\subsection{CGMF}

The study of CGMF with \delfin{} was limited to \isotope{Cf}{252} (fig.~\ref{fig:nubar-cf252sf}). 
Given the close affinity between the de-excitation model implemented in CGMF and the one used in TALYS (both based on the Hauser-Feshbach model), large discrepancies were not expected. CGMF is one of few codes that provides Z(A) distributions. The \delfin{} calculations were therefore performed twice, with the distribution taken from Wahl and CGMF, respectively.

Compared with the stand-alone code, \delfin{} consistently, though only slightly, overestimates  \nubarA. This is enough, however, to cause the total \nubar{} to differ by about 0.3 neutrons (table~\ref{tab:nubar}) from the tabulated value.
Around the peak, at A~=122, there is a significant difference between \delfin{}, stand-alone CGMF and data, and above mass 150 the \delfin{} result diverges from the stand-alone result in a similar manner as for GEF and FIFRELIN.

\subsection{FIFRELIN}

FIFRELIN, like TALYS and CGMF, also treats de-exci\-tation with the Hauser-Feshbach model. 
However, the excitation energy reported for each FF is the sum of both intrinsic and rotational energy~\cite{capote2016prompt}. As \delfin{} uses the whole \Eexc{} as energy available for de-excitation through neutron emission this leads to a general over-estimation of \nubarA{} over the whole mass range.
The only exception being in proximity of mass A~\num{= 132}, where the nuclei approach a spherical shape which reduces the available rotational energy.

Angular momentum ($J$) distributions from FIFRELIN were not available for this study, making it impossible to perform an exact subtraction of the rotational energy contribution, as done for GEF \cite{mattera2017methodology}.
However, following a description given by the developers~\cite{litaize2010investigation}, it was possible to reconstruct how $J$ is calculated and estimate the contribution from the rotational energy using
\begin{equation}
 E_{\mathrm{rot}} = \frac{\hbar^2 J (J+1)}{2 \Im }.
 \label{eqn:Jrms}
\end{equation}
Here $J$ is the total angular momentum and $\Im$ the moment of inertia. The latter was determined as \SI{50}{\percent} of the moment of inertia of a rigid spheroid \cite{litaize2010investigation}.

The result of this study is shown in figure~\ref{fig:nubar-u235+nth}~-~\ref{fig:nubar-cf252sf}. Once the rotational energy is subtracted using our simplified treatment, the agreement between the \nubarA{} obtained with \delfin{} and the one from FIFRELIN stand-alone improves significantly. FIFRELIN typically estimates a deeper minimum of the sawtooth and consequently put it at a too high mass ($A \approx 132$), compared to what is observed experimentally. Interestingly, this effect is enhanced when using \delfin{} for the de-excitation process. The same, but opposite, effect is also noted for the complementary fragments (around $A=A^{\mathrm{CN}}-132$) and for $A>150$.

%%% New paragraph from thesis %%%%%%

\subsection{\texttt{FREYA}}

\texttt{FREYA} is a Monte Carlo code that, like GEF, uses a simplified description of the de-excitation of the FFs and the behaviour of \nubarA{} obtained by the two models also show many similarities.
Using \delfin{} with \texttt{FREYA}, an interesting observation is that, for the case of  \isotope{U}{235}(n$_{\mathrm{th}}$,f), the relative number of neutrons emitted from the heavy fragments is larger than from the light side. This is in contradiction to what is obtained by the stand-alone \texttt{FREYA} and all the other codes, as well as experimental data (see Table~\ref{tab:nubar}). The behaviour is less evident for the \isotope{Pu}{239}(n$_{\mathrm{th}}$,f) reaction although the relative neutron emission from the heavy side increases when the de-excitation is handled by \delfin{}. This substantial difference between \delfin{} and \texttt{FREYA} can be explained by the introduction of an adjustable parameter $x$ in the \texttt{FREYA} code. This parameter is used to redistribute the \Eexc{} of the fragments to tune the results into agreement with experimental data on \nubarA{}~\cite{vogt2009event}.

Another trend observed in the case of \isotope{Pu}{239}(n$_{\mathrm{th}}$,f), from both the stand-alone version of \texttt{FREYA} and \delfin{}, is the steep decrease in \nubarA{} above mass $A= 150$. This behaviour is neither found in experimental data, nor in the data from any of the other codes.

It should be pointed out that the results presented here use distributions of \Eexc{} and \nubarA{} from an earlier version of the code~\cite{randrup2009calculation}. It would be interesting to repeat the same calculation with \Eexc{} distributions obtained from the more recent version, \texttt{FREYA 2.0}~\cite{verbeke2017}.

\subsection{PbP}

The results using \delfin{} with the PbP model for \isotope{U}{235}(n$_\mathrm{th}$,f) shows a striking difference in the height of the sawtooth with respect to the stand-alone code, as well as to the other models, especially for the light FFs and the total \nubar~(see table~\ref{tab:nubar}).
This is most likely due to the fact that the \Eexc($A$) at full acceleration includes rotational energy, that should not be available for neutron emission.
Since this model, unlike FIFRELIN, does not provide a detailed description for the estimation of \Erot{}, it was not possible to subtract this contribution (A. Tudora, in private communication).
It is interesting to note that PbP does not explicitly calculate \Erot{}, as it is a very important contribution to the non-statistical $\gamma$ emission. 

%%%%%%%%%%%%%%%%%%%%%%%%%%%%%%

\section{Summary and Outlook}

The computer code \delfin{} has been developed to handle de-excitation of fission fragments in TALYS, using excitation energies obtained from fission codes as input. Specifically, the code produces \nubarA{} distributions which can be compared with the results obtained with the stand-alone versions of the respective code and with experimental data. This makes it possible to test the assumptions going into the fission modelling, without the tuning to experimental data.

After validation of the methodology using the output from the GEF code, the code has been tested on five of the most commonly used models for the extraction of fission observables: \texttt{FREYA}, PbP, GEF, FIFRELIN and CGMF.

Codes based on the Hauser-Feshbach formalism (CGMF and FIFRELIN) seem to behave more consistently than the others.
However, the fact that TALYS uses the same model makes the comparison unfair.
Nevertheless, this is a verification that three implementations of the same model generally produce consistent results.
By not treating \Erot{} explicitly (like the PbP model), or by using a simplified model for the emission of neutrons in competition with $\gamma$ de-excitation (like \texttt{FREYA} and GEF), codes are disregarding phenomena that can be important for a correct reproduction of experimental data.

% Using \delfin{}, the \EexcA{} from FIFRELIN leads to a consistent over-estimation of the \nubarA{}.
% The excess \Eexc{} is compatible with the amount of rotational energy \Erot{} assigned by FIFRELIN to the fission fragments.
% Once the \Erot{} is subtracted from the energy available for neutron emission, \delfin{} reproduces quite well the results of FIFRELIN.
% This was expected, since the de-excitation in both cases is based on the same Hauser-Feshbach model.
% 
% The CGMF results show an interesting behaviour in the vicinity of mass A~\num{= 122}.
% While the \delfin{} calculation with the Z(A) from the Wahl systematics are in good agreement with the CGMF stand-alone calculation, when \delfin{} uses the Z(A) from the code itself, it over-estimates the \nubar.

Future developments of the \delfin{} code will involve the extraction of other quantities besides \nubarA.
For example, the possibility to look at $\gamma$ emission as well as prompt fission neutron spectra will be investigated.

\section*{Acknowledgement}
The authors are thankful to A.~Tudora, O.~Serot and P.~Talou for kindly providing the sets of excitation energies needed for this work.

This work was supported by the Swedish Radiation Safety Authority (SSM) and the Swedish Nuclear Fuel and Waste Management Co. (SKB).
%\end{acknowledgement}

%and \cite{RefJ}
%\subsection{Subsection title}
%\label{sec:2}
%as required. Don't forget to give each section
%and subsection a unique label (see Sect.~\ref{sec:1}).
%

%%%%%% For one-column wide figures use
%\begin{figure}
% Use the relevant command for your figure-insertion program
% to insert the figure file.
% For example, with the option graphics use
%\resizebox{0.75\textwidth}{!}{%
%  \includegraphics{leer.eps}
%}
% If not, use
%\vspace{5cm}       % Give the correct figure height in cm
%\caption{Please write your figure caption here}
%\label{fig:1}       % Give a unique label
%\end{figure}
%
%%%%%% For two-column wide figures use
%\begin{figure*}
% Use the relevant command for your figure-insertion program
% to insert the figure file. See example above.
% If not, use
%\vspace*{5cm}       % Give the correct figure height in cm
%\caption{Please write your figure caption here}
%\label{fig:2}       % Give a unique label
%\end{figure*}
%
%%%%%%%%%%%%% For tables use
%\begin{table}
%\caption{Please write your table caption here}
%\label{tab:1}       % Give a unique label
% For LaTeX tables use
%\begin{tabular}{lll}
%\hline\noalign{\smallskip}
%first & second & third  \\
%\noalign{\smallskip}\hline\noalign{\smallskip}
%number & number & number \\
%number & number & number \\
%\noalign{\smallskip}\hline
%\end{tabular}
% Or use
%\vspace*{5cm}  % with the correct table height
%\end{table}
%
% BibTeX users please use
\bibliography{mattera_Paper3}

\newcommand{\noopsort}[1]{}
\begin{thebibliography}{10}
\providecommand{\url}[1]{{#1}}
\providecommand{\urlprefix}{URL }
\expandafter\ifx\csname urlstyle\endcsname\relax
  \providecommand{\doi}[1]{DOI \discretionary{}{}{}#1}\else
  \providecommand{\doi}{DOI \discretionary{}{}{}\begingroup
  \urlstyle{rm}\Url}\fi

\bibitem{madland1982new}
D.G. Madland, J.R. Nix, Nucl. Sci. Eng. \textbf{81}(2), 213 (1982)

\bibitem{vladuca2001prompt}
G.~Vladuca, A.~Tudora, Ann. Nucl. Energy \textbf{28}(16), 1643 (2001)

\bibitem{tudora2006experimental}
A.~Tudora, Ann. Nucl. Energy \textbf{33}(11), 1030 (2006)

\bibitem{schmidt2010general}
K.H. Schmidt, B.~Jurado, Rapport technique, CENBG, CNRS/IN2P3  (2010)

\bibitem{randrup2009calculation}
J.~Randrup, R.~Vogt, Phys. Rev. C \textbf{80}(2), 024601 (2009)

\bibitem{vogt2009event}
R.~Vogt, J.~Randrup, J.~Pruet, W.~Younes, Phys. Rev. C \textbf{80}(4), 044611
  (2009)

\bibitem{kawano2010monte}
T.~Kawano, P.~Talou, M.B. Chadwick, T.~Watanabe, J. Nucl. Sci. Technol.
  \textbf{47}(5), 462 (2010)

\bibitem{talou2011advanced}
P.~Talou, B.~Becker, T.~Kawano, M.~Chadwick, Y.~Danon, Phys. Rev. C
  \textbf{83}(6), 064612 (2011)

\bibitem{becker2013monte}
B.~Becker, P.~Talou, T.~Kawano, Y.~Danon, I.~Stetcu, Phys. Rev. C
  \textbf{87}(1), 014617 (2013)

\bibitem{fifrelin}
O.~Litaize, O.~Serot, L.~Berge, Eur. Phys. J. A \textbf{51}(12), 1 (2015)

\bibitem{litaize2010investigation}
O.~Litaize, O.~Serot, Phys. Rev. C \textbf{82}, 054616 (2010)

\bibitem{schmidt2014general}
K.H. Schmidt, B.~Jurado, C.~Amouroux, JEFF Report \textbf{24} (2014)

\bibitem{tudora2015comparing}
A.~Tudora, F.J. Hambsch, I.~Visan, G.~Giubega, Nucl. Phys. A \textbf{940}, 242
  (2015)

\bibitem{andreyev2017nuclear}
A.~Andreyev, K.~Nishio, K.H. Schmidt, Rep. Prog. Phys. \textbf{81}, 016301
  (2017)

\bibitem{tudora2017comprehensive}
A.~Tudora, F.J. Hambsch, Eur. Phys. J. A \textbf{53}(8), 159 (2017)

\bibitem{talys}
A.J. Koning, S.~Hilaire, M.C. Duijvestijn, in \emph{International Conference on
  Nuclear Data for Science and Technology} (EDP Sciences, 2007), pp. 211--214

\bibitem{mattera2017methodology}
A.~Mattera, A.~Al-Adili, M.~Lantz, S.~Pomp, V.~Rakopoulos, A.~Solders, in
  \emph{EPJ Web of Conferences}, vol. 146 (EDP Sciences, 2017), vol. 146, p.
  04047

\bibitem{wahl1988nuclear}
A.C. Wahl, Atom. Data Nucl. Data \textbf{39}(1), 1 (1988)

\bibitem{matteraPhDthesis}
A.~Mattera, Studying neutron-induced fission at {IGISOL}-4: From neutron source
  to yield measurements and model comparisons.
\newblock Ph.D. thesis, Acta Universitatis Upsaliensis (2017)

\bibitem{geltenbort1986precision}
P.~Geltenbort, F.~G{\"o}nnenwein, A.~Oed, Radiat. Eff. Defect. S.
  \textbf{93}(1-4), 57 (1986)

\bibitem{gook2014cf252}
A.~G{\"o}{\"o}k, F.J. Hambsch, M.~Vidali, Phys. Rev. C \textbf{90}(6), 064611
  (2014)

\bibitem{endf}
M.B. Chadwick, M.~Herman, P.~Oblo{\v{z}}insk{\`y}, et~al., Nucl. Data Sheets
  \textbf{112}(12), 2887 (2011)

\bibitem{PhysRevC.98.044615}
A.~G\"o\"ok, F.J. Hambsch, S.~Oberstedt, M.~Vidali, Phys. Rev. C \textbf{98},
  044615 (2018)

\bibitem{schmidt2010energysorting}
K.H. Schmidt, B.~Jurado, Phys. Rev. Lett. \textbf{104}, 212501 (2010)

\bibitem{capote2016prompt}
R.~Capote, et~al., Nucl. Data Sheets \textbf{131}(LA-UR-15-26188) (2016)

\bibitem{verbeke2017}
J.M. Verbeke, J.~Randrup, R.~Vogt, Freya 2.0: New developments in fission chain
  modeling.
\newblock Tech. Rep. PROC-732580, Lawrence Livermore National Laboratory
  (LLNL), Livermore, CA (2017)

\end{thebibliography}
\bibliographystyle{spphys.bst}
%
% Non-BibTeX users please use
%\begin{thebibliography}{}
%
% and use \bibitem to create references.
%
%\bibitem{RefJ}
% Format for Journal Reference
%Author, Journal \textbf{Volume}, (year) page numbers.
% Format for books
%\bibitem{RefB}
%Author, \textit{Book title} (Publisher, place year) page numbers
% etc
%\end{thebibliography}

\end{document}